\documentstyle[12pt]{article}

\baselineskip=\normalbaselineskip

\begin{document}

\begin{titlepage}

\begin{flushright}
YITP-95/KUCP-0082  \\ 
September 1995     
\end{flushright}
\vskip 1cm

\begin{center}
{\Large  
Wavelet Analysis \\
of \\
One-Dimensional Cosmological Density Fluctuations
}\vskip 2cm

{\large  
Yoshi Fujiwara
}\vskip 0.7cm

{\sl  
Yukawa Institute for Theoretical Physics,\\
Kyoto University, Kyoto 606, Japan}
\vskip 1.5cm

{\large  
Jiro Soda
}\vskip 0.7cm

{\sl  
Department of Fundamental Sciences,
FIHS,\\
Kyoto University, Kyoto 606, Japan}
\end{center}
\vskip 2cm

\begin{abstract}
Wavelet analysis is proposed as a new tool for studying the
large-scale structure formation of the universe. To reveal its
usefulness, the wavelet decomposition of one-dimensional cosmological
density fluctuations is performed. In contrast with the Fourier
analysis, the wavelet analysis has advantage of its ability to keep
the information for location of local density peaks in addition to
that for their scales. The wavelet decomposition of evolving density
fluctuations with various initial conditions is examined. By comparing
the wavelet analysis with the usual Fourier analysis, we conclude that
the wavelet analysis is promising as the data analysis method for the
Sloan Digital Sky Survey and COBE.
\end{abstract}

\end{titlepage}

\setlength{\normalbaselineskip}{20pt plus 0.2pt minus 01.pt}
\baselineskip=\normalbaselineskip

\section{Introduction}

Recent large-scale sky survey (e.g. CfA) reveals the fertile
large-scale structure of the universe \cite{Peebles}. It is natural to
consider it as a consequence of gravitational instability from
small-amplitude primordial fluctuations. The origin of the primordial
fluctuations is usually ascribed to the quantum fluctuations around
the Planck time which are stretched to a scale larger than the Hubble
horizon during the inflationary stage and classicalize due to
decoherence.  Although the above picture is yet rather speculative, it
encourages us to explain the large scale structure of the present
universe from the physical point of view.

Based on the cosmological principle, the fairness of the observational
data is assumed and the results are statistically interpreted. Then it
is important to determine the statistical properties of the various
cosmic fields that can be used to describe the matter distribution and
motion in the expanding universe. For this purpose, the correlation
functions are useful, indeed, two-point correlation function is well
studied observationally which yields the fractal picture of the
universe \cite{Peebles}. Other methods, such as the power spectrum,
the topology of the iso-density surface \cite{GMD} etc. are also used.
As the inflation theory naturally predicts the random Gaussian initial
fluctuations, the power spectrum analysis is useful as far as the
density field stays in the linear regime. However, the non-linear
dynamics causes mode-mode coupling which produces non-Gaussianity
generating reduced $n$-point correlation functions. In the terminology
of the Fourier analysis, such nonlinear effect is described by the
phase correlation. It was found in \cite{SutoSoda} that it is
difficult to obtain information of nonlinear effect in this way.

Alternatively, one can works in real space, i.e., N-body simulation
\cite{Suto}. However, in real space, the information of the scale
(wavenumber) is lacked. Thus, the Fourier spectrum analysis and the
real space analysis is complementary. This situation is analogous to
quantum mechanics. In quantum theory, the Wigner function which is
defined on the phase space is known to be useful.  Instead, in this
paper, we propose to use the wavelet analysis \cite{Wavelet} to
characterize the density perturbation. In contrast with the Fourier
analysis, localized basis functions are used in the wavelet
analysis. Therefore, the information of the phase, or the position, is
stored explicitly. In addition to the information of location, the
wavelet analysis gives the scale information, which is similar to
Wigner function in quantum mechanics. Another advantage of the wavelet
analysis should be stressed, that is, it can be performed with the
data in certain compact region, in contrast the Fourier analysis needs
the information of the whole domain.

To demonstrate its usefulness, we shall use one-dimensional
cosmological model for which Zeldovich exact solution is known
\cite{Zel}.  There are various ways to choose the basis functions in
the wavelet analysis, however, as we concentrate on the non-linear
effects it is sufficient to use an arbitrary basis, for which we
choose the so-called Spline 4 wavelet. It is important to specify the
useful statistics based on the wavelet analysis and calculate it from
the physics and compare it with observation. To accomplish this task,
thorough understanding of the non-linear dynamics based on the wavelet
analysis is required. It is our present work that attack to this
problem, as a first step in this direction. Our final goal is the
application to the DSS project.

The plan of the paper is the following. In Sec.2, we review the
wavelet analysis briefly. One-dimensional model is explained in Sec.3.
The numerical results are presented in Sec.4. We have concluded with
the summary and discussion of future problems in Sec.5.

\section{Wavelet Analysis}

Fourier transform method is a useful tool in data analysis since it
enables us to decompose data into components with different
scales. Many fundamental properties of physical systems have been
described in terms of Fourier spectrum, that is, the amplitude of
Fourier coefficients. However, since Fourier spectrum ignores the
phase of each Fourier coefficient, it lacks information about
positions of local events which are difficult to observe from the
characteristics of the spectrum. The Fourier spectrum analysis
therefore encounters difficulty in analyzing data which include
different kinds of local structures.

Discrete wavelet analysis is invented to circumvent this
difficulty. In the Fourier analysis, the basis functions of expansion
are the familiar sines and cosines. In the wavelet analysis, the basis
functions are somewhat more complicated localized functions, the
so-called wavelet. Hence, it can detect both the location and the
scale of the structure.

Let us review the wavelet analysis briefly. The reader is referred to
the reference \cite{Wavelet} for complete information. The essential
ingredient of the wavelet analysis is the scaling functions $\phi$
which satisfy the two-scale relation:
\begin{equation}
 \phi(x) = \sum_{k \in Z} p_k \phi (2x -k) \ ,
\end{equation}
where $ \{ p_k \}$ is called two-scale sequence. The basis functions
of Fourier analysis are $\exp ikx$ which can be regarded as dilation
of $\exp ix$. In the wavelet analysis, the mother wavelet is
localized, hence to cover the whole range it is necessary to perform
the translation in addition to dilation. Discrete wavelet analysis
uses $2^j$ dilation and the $k/2^j$ translation. Then, $\phi (2^j
(x-k/2^j))=\phi (2^j x -k)$. The index $j$ denotes the level of the
resolution and $k$ represents the location. Let $V_j$ be the space
spanned by the scaling functions of level $j$, $\{ \phi(2^j x -k)
\}_{k\in Z}$. As the two-scale relation yields $\phi (2^j x)= \sum_k
p_k \phi(2^{j+1} x -k) $, one sees $V_j \subset V_{j+1}$. Thus once
the scaling functions are given, the hierarchical structure,
called multiresolution analysis, is generated as
\begin{equation}
 \cdots \subset V_{j-1} \subset V_j \subset V_{j+1} \subset \cdots \ .
\end{equation}
Similarly, the mother wavelet is defined by the following two-scale
relation:
\begin{equation}
 \psi(x) = \sum_{k \in Z} q_k \phi (2x -k) \ .
\end{equation}
Here the sequence $\{q_k \}$ is considered to define the mother
wavelet.  Let $W_j$ be the space spanned by the wavelet functions of
level $j$, $ \{ \psi (2^j x -k) \}_{k\in Z} $. Now, an arbitrary
function $f_j \in V_j$ can be expressed as
\begin{equation}
   f_j (x) = \sum_k  c_k^{(j)} \phi (2^j x -k) \ ,
\end{equation}
where $c_k^{(j)}$ is the expansion coefficients.  The multiresolution
analysis suggests the direct sum decomposition,
\begin{equation}
    V_j = V_{j-1} \bigoplus W_{j-1} \ .
\end{equation}
The key fact is the uniqueness of this decomposition in the form
\begin{equation}
  f_j = f_{j-1}  + g_{j-1} \  ,
\end{equation}
where
\begin{equation}
  g_j (x) = \sum_k d_k^{(j)} \psi (2^j x -k) \ .
\end{equation}
Here, the expansion coefficients $d_k^{(j)}$ is the correspondent one
to the Fourier coefficients. The major difference is the number of the
index, that is, in the wavelet analysis there are two indices which
indicate the location and the scale. The repeated decomposition gives
\begin{equation}
  f_j = g_{j-1} + g_{j-2} + g_{j-3} + \cdots  \ .
\end{equation}
The above decomposition is the wavelet decomposition analysis.

If the data is given at $2^j$ points, one can construct interpolating
function. This function is regarded as $f_0$. Then the above
decomposition procedure gives
\begin{equation}
  f_0 = g_{-1} + g_{-2} + \cdots + g_{-j} \ .
\end{equation}

Of course there are many wavelets that have local support.  We use the
well-known spline wavelet of order 4 and Daubechies wavelet of order 4
as typical wavelets. We do not give their details; instead we show
their plots in Fig.~1 and Fig.~2.

\section{Model and Formalism}

Throughout the present analysis, we adopt a matter dominated,
spatially flat universe, i.e. the Einstein-deSitter model, for
simplicity. Furthermore, we consider one-dimensional self-gravitating
system by imposing plane symmetry. Let us consider only the
situation in which the Newtonian approximation is valid. Then the
basic equations are given by
\begin{eqnarray}
   {\partial \over \partial t}\delta (x,t)
  +{1\over a(t)}{\partial \over \partial x}
   [ v(x,t) ( 1 + \delta (x,t) ) ] &=& 0 \ ,       \\
   {\partial \over \partial t} v(x,t)
  +{1\over a(t)} v(x,t){\partial \over \partial x} v(x,t)
  +{\dot a \over a} v(x,t)
  + {1\over a(t)} {\partial \over \partial x}\phi (x,t) &=&   0    \ ,     \\
   {\partial^2 \over \partial x^2} \phi    =
   {3\over 2} ({\dot a \over a})^2 a^2 \delta(x,t) & &  \ ,
\end{eqnarray}
where $\delta$ and $v$ are the density perturbation and the peculiar
velocity field, respectively and $x$ is the comoving coordinate. Since
we assume the flat universe, the scale factor $a(t)$ is proportional
to $t^{2/3}$.

It is well known that there is an exact solution in a fully non-linear
field, i.e. Zeldovich solution \cite{Zel}. The basic idea underlying
the Zeldovich solution is the transformation from Eulerian, $x$, to
Lagrangian, $q$, space coordinates:
\begin{equation}
  x = q + B(t) S(q) \ ,
\label{xq}
\end{equation}
where $B(t)$ is a function of time to be determined later and $S(q)$
is an arbitrary function of $q$. Then the density fluctuation is
explicitly given by
\begin{equation}
   \delta (x,t) = - {B(t) S^{\prime} (q) \over 1 + B(t) S^{\prime} (q) } \ ,
\label{delta}
\end{equation}
where a prime denotes the derivative with respect to $q$.  One can
easily verify that the basic equations are solved automatically
provided that $B(t)$ satisfies
\begin{equation}
 {d^2 \over dt^2} B(t) + 2{\dot a \over a} {d \over dt} B(t)
      = {3\over 2} ({\dot a \over a})^2 B(t) \ ,
\label{funcb}
\end{equation}
where a dot denotes a differentiation with respect to $t$.

The above equation is nothing but the same equation for the growth
factor of pressureless matter in linear perturbation theory, and
consists of a decaying and a growing solution. In what follows, we
take the growing solution and use $B(t)$ as a time variable instead
of $t$.  In the linear stage, the density perturbation becomes
\begin{equation}
  \delta (x,t) \sim -B(t) S^{\prime} (x) \ ,
\label{linear}
\end{equation}
hence the initial condition is determined by giving the arbitrary
function $S(q)$. Also we set $B_{\rm init} =1$. Let us consider the
following form
\begin{equation}
  S(q) = \sum_{k=1}^{k_c }\epsilon_n k^{n-1}\sin(kq +\alpha_k) \  ,
\label{funcs}
\end{equation}
where $\epsilon_n$ is independent of $k$, the power spectrum of the
initial density fluctuations, $P(k)$, is essentially characterized by
the spectral index $2n$ as $P(k)\propto k^{2n}$.

\section{Numerical Results}

We performed wavelet analysis for the Zeldovich solution of density
perturbation in one-dimension. The numerical analysis was done for
different initial power spectra given by $P(k)\propto k^{2n}$ and with
the use of two mother wavelets. In this section, we show the numerical
results for power spectra of $2n=2,0,-2$. Spline wavelet is mainly
used in the analysis, while we show Daubechies wavelet analysis only
for comparison.

We impose periodic boundary condition on a range $x\in[0,2\pi]$ in
one-dimension. The range is divided into evenly spaced $2^9=512$ small
intervals. At each time $t$, the profile of density fluctuation
$\delta(t,x)$, given by the solution (\ref{delta}), is sampled at each
intervals. Then the sampled data is transformed by wavelet transform
using spline 4 wavelet. In this wavelet analysis, $\delta(t,x)$ can be
decomposed into 9 levels, as is determined by the above
discretization. The sampled data is at the same time transformed by
Fourier transform so as to obtain power spectrum at each time. The
time evolution is followed up to nonlinear stage, but well before the
appearance of first orbit-crossing which occurs when the
transformation (\ref{xq}) becomes singular so that our fluid
approximation breaks down.

Initial profile of density contrast is given by the function $S(q)$ in
(\ref{funcs}), where $\alpha_k$ ($k=1,\cdots,k_c$) are randomly given
and we choose $k_c=10$. The overall constant $\epsilon_n$ is chosen so
that the initial amplitude of $\delta$ have maximum about 0.01. To be
explicit, we tabulate them:
  \begin{displaymath}
  \begin{array}{cc}
  2n & \epsilon_n  \\ \hline
  2  & 0.0004  \\
  0  & 0.002  \\
  -2 & 0.04
  \end{array}
  \end{displaymath}

We first show the wavelet analysis for linear growth of density
fluctuation, given by (\ref{xq}), so as to compare it with that for
nonlinear evolution. Fig.~3~(a) depicts the time evolution of linear
growth for $P(k)\propto k^2$, extrapolated to quasi-nonlinear
stage. Fig.~3~(b)(c) show the wavelet analysis for the initial and
late-time profile. ({\it Remark\/}: Our plot of the wavelet analysis,
like Fig.~3~(b), is such that the relative magnitude of the amplitudes
in 9 levels is meaningful. But the absolute magnitude is different
from plot to plot. This remark applies to every plots of wavelet
analysis below.) Corresponding to about 8 bumps of the density
fluctuation we have signals at levels $-5$ and $-6$. In this linear
growth of density fluctuation, we see no qualitative change of signals
at each levels in the wavelet analysis (b) and (c).  The magnitudes of
signals in each levels simply increase in accordance with the linear
growth of $\delta(t,x)$.

Fig.~4~(a)--(f) show the same case $P(k)\propto k^2$ for the identical
initial condition as above. Fig.~4~(a) is the time evolution of
density fluctuation up to full nonlinear stage. We wavelet-analyzed it
for four stages --- linear, quasi-linear, nonlinear and highly
nonlinear stages and show the results in the sequence of
Fig.~4~(c)--(f). In contrast with the linear case above, we can
observe that there is transfer between levels. Indeed, the level
$j=-4$ signal slightly appear from linear (c) to quasi-nonlinear stage
(d). As the density fluctuation increases, higher levels become to
have signals as is seen from (e) and (f). This \lq\lq power-transfer
between levels\rq\rq\ clearly represents the mode-mode coupling in
quasi- and full-nonlinear growth.

Moreover, because density peak becomes more and more narrow due to
gravitational effect, each peak makes a signal at the corresponding
locus and in higher and higher levels during the time evolution. The
wavelet analysis explicitly shows this situation in these figures.  It
is one of the advantages for the wavelet analysis that a kind of
position-wavenumber representation is possible and yields enough
information of the density peaks in real space. In the standard
Fourier analysis, if one knows the power spectrum, one can in
principle obtain such information from the phase correlation, but it
is very inefficient to recover the real space structure in that
way. In Fig.~4~(b), we show the power spectra for the same time
evolution. This result is in agreement with the general tendency that
the power-transfer occurs so that the power density tends to become
equal among all modes \cite{SutoSoda}.

We performed similar analyses for initial power spectra $P(k)\propto
k^0$ and $k^{-2}$. The results are shown in Fig.~5~(a)--(f) and
Fig.~6~(a)--(f) respectively. The \lq\lq power-transfer between
levels\rq\rq\ is observed also in these cases.

In Fig.~5, additional feature can be found concerning a local
structure of density peak. A bump present in the initial condition
grows a strong peak in the nonlinear stage. (See the first peak in
Fig.~5~(a).) As is found from the wavelet analyses for the late-time
behavior in Fig.~5~(e) and (f), this strong density-peak yields a
characteristic series of localized signals in the levels $j=-2,-3,-4$.
This feature corresponds to a self-similar evolution of the density
peak. In terms of the power-spectra in Fig.~5~(b), this corresponds to
a power-law at high wavenumber appearing in the late-time spectrum
there. This point has been studied in \cite{Zel} \cite{GoudaNaka},
where it was shown that independently of initial conditions
$|\delta_k|$ becomes to obey $k^{-1/3}$ power-law asymptotically. This
universal behavior occurs prior but rather closely to the
orbit-crossing singularity. However, this feature may be uninteresting
in the viewpoint of large-scale distribution of galaxies, in which one
is concerned with some \lq\lq average\rq\rq\ density field rather than
the appearance of singularity.

In Fig.~6, while a similar feature appears more prominently in (e) and
(f), we have still another interesting observation. For this initial
condition, a rather large-scale void is formed where $\delta$ is
negative. Since its wavelength is almost $2\pi$, the most
coarse-graining level $-9$ has the corresponding signal in the wavelet
analyses (c) and (d). The signal remains to be observed in spite of
the presence of the strong density peak that was mentioned above. (See
(e) and (f).) Wavelet analysis is also well adapted for the study of
such global structures as well as the local ones.

The last result that we present is for comparison between the spline 4
wavelet analysis and that by Daubechies 4 wavelet. Fig.~7~(a)--(d)
show the Daubechies wavelet analysis for the initial power spectrum
$P(k)\propto k^2$ with the identical initial condition as the above
Fig.~4. It can be seen from the result that signals in each levels are
more distinctive than in the spline wavelet, so the Daubechies wavelet
is sensitive in analyzing the local structure. The basic feature is
not different between the two wavelet analyses.

\section{Conclusion and Discussions}

In order to understand the large scale structure of the universe, it
is important to see what aspects we should look at. The power spectrum
analysis is used so far. In the near future, the Sloan digital sky
survey project (DSS) will begin and make a map of the whole
universe. Hence, more elaborative tool of analysis is necessary to
characterize the huge data. In this paper, we proposed the wavelet
analysis as a new tool. In order to study its usefulness, we utilized
the one-dimensional cosmological model and examined the non-linear
dynamics from the point of view of the wavelet analysis. In the
calculation of the dynamical evolution, we made use of the Zeldovich
exact solution. First of all, the results of the linear evolution
indicates no power-transfer between levels. In the full non-linear
analysis, three type of the initial conditions are prepared. Due to
non-linear interactions, higher level structures are generated in the
location of the peaks. For comparison, we also presented the power
spectrum evolution.  We found that formation of local strong
density-peak can be studied both from the real and \lq\lq level\rq\rq\
spaces in the wavelet analysis. This is based on its mathematical
scheme that one can locally zoom-in and zoom-out a data. Furthermore,
for completeness, we presented the analysis using different mother
wavelet.

If one looks at the power spectrum, the initially prepared mode does
not grow in the non-linear region as was studied previously (see
\cite{SutoSasaki} for example). On the other hand, in the real space,
the density fluctuations will unboundedly grow in some spatially
compact region due to nonlinear effect. This apparent contradiction is
clearly a hasty consequence of the Fourier analysis and reveals a
disadvantage of the method. The underlying reason of this misleading
argument is rather trivial, that is, the power-transfer accumulatively
generates high frequency modes of small amplitude. Thus, in order to
follow the system that evolves into the nonlinear stage, it is
necessary to know the information about which modes are newly
generated and how they form a structure in the real space at the same
time. Clearly the Fourier analysis is not adequate for doing it. It is
wavelet analysis that can detect the location (phase) of the newly
generated structure. In the wavelet analysis, it is easy to see that
accumulation of these structures is compensating the slowing down the
growth of the original mode. Thus, it turned out that the wavelet
analysis is very useful to understand the non-linear mode-mode
interaction.

It should be stressed, however, that we regard the wavelet and Fourier
analyses as complementary to each other. The power spectrum in Fourier
analysis is an important tool that bridges theoretical prediction and
the observational data. By using an orthogonal wavelet, such as
Daubechies wavelet, it is possible to define a power spectrum in terms
of the wavelet and to relate it with that in Fourier analysis. We may
use the former to argue a kind of \lq\lq local spectra\rq\rq\ for
positions of different density contrast. (Indeed, though in different
field, recent study of turbulence in fluid mechanics found such a
concept useful \cite{YamadaOhki}.) Since the wavelet analysis is
entirely based on a similarity of the two-scale relation as was
explained, it would be well adapted for the study of gravitational
system where local and similar evolution is considered to be
important. Statistical average of such local spectra would yield a new
statistical description of the large-scale structure of the
universe. Theoretical models could be analyzed in terms of the wavelet
and be compared with such detailed statistical quantities. In a
separate paper, we will report the investigation of these statistics.

As other future tasks, the extension to two- or three-dimensional
space is planed. In addition to the DSS, it is interesting to apply
the wavelet analysis to COBE data. Because the wavelet analysis does
not require the whole sky data, the galactic plane does not cause any
trouble in contrast to the spherical harmonics analysis. What is more
important is how to compare such observational data obtained by the
wavelet analysis with physical theories and models. Our strategy
is the following; first, we will define the useful statistics such as
the local power spectrum.  It gives more information than the usual
power spectrum.  Next, the local power spectrum is measured from the
observation.  On the other hand, we can calculate that quantity from
the physical theory like as the inflation theory. Comparing both
quantities will give powerful constraints on the theoretical
parameters. Thus we believe that the wavelet analysis is quite
promising and will open a new way for understanding the large scale
structure of the universe.

\section*{Acknowledgements}

We would like to express our thanks to Susumu Sakakibara for allowing
us to implement his {\sl Mathematica\/} (Wolfram Inc.) program in our
numerical algorithm. The work is supported by Monbusho Grant-in-Aid
for Scientific Research No.~07740216.



\newpage
\pagestyle{empty}

\begin{center}
{\Large Figure Captions}
\end{center}

\begin{description}
\item[Fig. 1] Spline 4 wavelet.
\item[Fig. 2] Daubechies 4 wavelet.
\item[]
\item[Fig. 3 (a)] Time evolution of linear growth of density
  fluctuation for initial power spectrum with $2n=2$ at
  $B(t)=10,45,74.2$.
\item[Fig. 3 (b)] Wavelet decomposition analysis for the initial
  profile at $B(t)=1$.
\item[Fig. 3 (c)] Wavelet decomposition analysis for the late-time
  profile at $B(t)=74.2$.
\item[]
\item[Fig. 4 (a)] Time evolution of density fluctuation for initial
  power spectrum with $2n=2$ at $B(t)=10,27.3,45,54.9$.
\item[Fig. 4 (b)] Power spectra at the corresponding times.
\item[Fig. 4 (c)] Wavelet decomposition analysis at $B(t)=10$.
\item[Fig. 4 (d)] Wavelet decomposition analysis at $B(t)=27.3$.
\item[Fig. 4 (e)] Wavelet decomposition analysis at $B(t)=45$.
\item[Fig. 4 (f)] Wavelet decomposition analysis at $B(t)=54.9$.
\item[]
\item[Fig. 5 (a)] Time evolution of density fluctuation for initial
  power spectrum with $2n=0$ at $B(t)=10,27.3,60.7,74.2$.
\item[Fig. 5 (b)] Power spectra at the corresponding times.
\item[Fig. 5 (c)] Wavelet decomposition analysis at $B(t)=10$.
\item[Fig. 5 (d)] Wavelet decomposition analysis at $B(t)=27.3$.
\item[Fig. 5 (e)] Wavelet decomposition analysis at $B(t)=60.7$.
\item[Fig. 5 (f)] Wavelet decomposition analysis at $B(t)=74.2$.
\item[]
\item[Fig. 6 (a)] Time evolution of density fluctuation for initial
  power spectrum with $2n=-2$ at $B(t)=1.5,3.8,8.3,10$.
\item[Fig. 6 (b)] Power spectra at the corresponding times.
\item[Fig. 6 (c)] Wavelet decomposition analysis at $B(t)=1.5$.
\item[Fig. 6 (d)] Wavelet decomposition analysis at $B(t)=3.8$.
\item[Fig. 6 (e)] Wavelet decomposition analysis at $B(t)=8.3$.
\item[Fig. 6 (f)] Wavelet decomposition analysis at $B(t)=10$.
\item[]
\item[Fig. 7 (a)] Daubechies analysis for the same time evolution as
  in Fig.~4. This is at $B(t)=10$ corresponding to Fig.~4~(c).
\item[Fig. 7 (b)] The same as (a) and at $B(t)=27.3$ corresponding to
  Fig.~4~(d).
\item[Fig. 7 (c)] The same as (a) and at $B(t)=45$ corresponding to
  Fig.~4~(e).
\item[Fig. 7 (d)] The same as (a) and at $B(t)=54.9$ corresponding to
  Fig~.4~(f).

\end{description}

\end{document}